# Selection of strain and fitting schemes for calculating higher-order elastic constants

Mingqing Liao[1,2], Yong Liu[1,3], Fei Zhou[1,4], Tianyi Han[1], Danni Yang[1], Nan Qu[1], Zhonghong Lai[1,5], Zi-Kui Liu[2], Jingchuan Zhu[1,3, *]

[1]School of Materials Science and Engineering, Harbin Institute of Technology, Harbin, Heilongjiang, 150001, China

[2]Department of Materials Science and Engineering, The Pennsylvania State University, University Park, PA, 16802, USA

[3]National Key Laboratory for Precision Hot Processing of Metals, Harbin Institute of Technology, Harbin 150001, Heilongjiang, China

[4]MIIT Key Laboratory of Critical Materials Technology for New Energy Conversion and Storage, School of Chemistry and Chemical Engineering, Harbin Institute of Technology, Harbin, Heilongjiang, 150001, China

[5]The Centre of Analysis Measurement, Harbin Institute of Technology, Harbin, Heilongjiang, 150001, China

**Abstract:** Criteria of selecting strain and fitting schemes are proposed for the calculation of higher-order elastic constants more efficiently, robustly and accurately. As demonstrated by the third-order elastic constants (TOECs) of diamond, the proposed method is 3-5 times faster than existing methods, and the range of strain for getting correct TOECs is expanded. In addition, our result provides an evidence for the inaccuracy of some previous experiments caused by higher-order effect, and the difference among experiments and several different theoretical methods is resolved. Finally, we give the recommend TOECs values for diamond.

* Corresponding author: fgms@hit.edu.cn (Jingchuan Zhu)

Higher order elastic constants (HOECs) describe the nonlinear elastic response of materials. Based on the HOECs, many anharmonic properties can be explored, such as thermal expansion and Grüneisen parameters [1], temperature and pressure dependence of the second order elastic constant (SOECs) [2–4], ideal strength and ductility [5]. However, it is still a challenge to obtain the accurate HOECs efficiently.

To date, the experimental measurement of the HOECs remains difficult with high uncertainty, especially for those materials with extreme properties [6,7]. Computationally, there are mainly two methods used in the literature to calculate the HOECs, namely lattice dynamical theory and homogeneous deformation method [8]. In lattice dynamical theory, the force constants of the neighbor or next-neighbor bonds are required, and it is often combined with molecular mechanics and molecular dynamics simulations and applied to covalent [9,10] and ionic [11] crystals. The homogeneous deformation method is much simple because there is no need to consider the complex inner-interaction of crystal, and thus has been widely applied to calculate TOECs [12,13] in combination with first-principles calculations. The homogeneous deformation method can be divided into two classes, strain-energy method (SEM) and strain-stress method (SSM). The strain-energy method has been well-developed. Zhao et al. [14] deduced the method of TOECs for arbitrary symmetry, which was expanded to forth-order elastic constants by Wang et al. [15] and to 2D materials by Cadelano et al. [16]. In past few years, this method attracted many applications [1,4,7,17]. The strain-stress method was firstly introduced by Nielsen et al. [12,13] where three strains ([001], [110] and [111]) were applied to cubic crystal. Subsequently, Hmiel et al. [7] developed the longitudinal stress-uniaxial strain approach (LSUS) to match the result got by shock wave experiments [18]. Recently, Jong et al. [5] used 21 different strain modes containing all 56 unique TOECs to evaluate the second derivative of Piola-Kirchhoff tensor. Meanwhile, Cao et al. [19] separated each SOECs and TOECs by the numerical differentiation (SEP-ND) of the second Piola-Kirchhoff tensor and extend it to arbitrary symmetry.

Overall, there exist two challenges in calculating HOECs. The first one is the efficiency: the existing methods [5,7,19] are not efficient enough, causing it difficult to calculate the low symmetry structure. The second one is the robustness: the HOECs is quite sensitive to strain amplitude [14] and the results between different theoretic methods disagree with each other, as well as with experiments.

Diamond is one important example. The results by SEM [7,20] agree well with SEP-ND [19], but there exists a huge gap when compared with the longitudinal stress-uniaxial strain approach (LSUS) [7] and experiments [18,21].

Here we introduce a set of criteria on selecting strain modes and fitting strategies to make the calculation of HOECs more efficient and more accurate using the TOECs of diamond as an example. We demonstrate the fundamental understanding of the disagreement among different methods and the approach to resolve the differences. All of the first-principles calculations based on the density functional theory (DFT) are performed using CASTEP [22] with ultra-soft pseudopotential [23], and Perdew-Burke-Ernzerhof (PBE) functional in generalized gradient approximation (GGA) [24] framework. The cutoff energy and k-points are determined by convergence test as 400 eV and 8×8×8, respectively. To eliminate the error caused by different symmetry [19], a tiny (~$10^{-10}$ lattice constants) random perturbation on atomic position is added to the initial structure. The cell is fixed for the deformed structure. The convergence criterions of relaxation for deformed-structure are $5\times10^{-6}$ eV/atom in energy, 0.01 eV/Å in residual force per atom, $5\times10^{-4}$ Å in max displacement and 0.05 GPa in max stress. The workflow of calculating TOECs is presented in Elastic3rd code [25].

The base of all branches of homogeneous deformation method is continuum elasticity theory. The energy of crystal can be expressed as a function of Lagrangian strain (**η**) and the higher-order derivation of energy with respect to **η**, which defines the HOECs [26]. The Lagrangian stress ($t_{ij}$) is defined as the derivative of energy on **η** [12], and can be expressed as follows [13].

$$t_\alpha = t_{ij} = \sum_{k,l=1}^{3} C_{ijkl}\eta_{kl} + \frac{1}{2!}\sum_{k,l,m,n=1}^{3} C_{ijklmn}\eta_{kl}\eta_{mn} + \cdots = \sum_{\beta=1}^{6} B_2^\beta C_{\alpha\beta}\eta + \frac{1}{2!}\sum_{\beta,\gamma=1}^{6} B_3^{\beta,\gamma} C_{\alpha\beta\gamma}\eta^2 + \cdots = A_2\eta + \frac{1}{2!}A_3\eta^2 + \cdots \quad (1)$$

where the subscript of $i$, $j$, $k$, $l$, $m$ and $n$ are the tensor index, $\alpha$, $\beta$ and $\gamma$ are Voigt's notation. $C_{ijkl}$ ($C_{\alpha\beta}$) and $C_{ijklmn}$ ($C_{\alpha\beta\gamma}$) the second- and third-order elastic constants, and $B_2^\beta$ and $B_3^{\beta,\gamma}$ constants determined by **η**. $A_2$ and $A_3$ are the linear combination of SOECs and TOECs, respectively.

In **Eq.(1)**, the Lagrangian strain-stress relationship can be calculated by first-principles code for a given **η**, which determines the value of $A_2$ and $A_3$. (ref. [27] for details) The elastic constants are obtained by solving linear equation group about independent SOECs and TOECs. In the process, two things are variable. One is the method to get the value of $A_2$ and $A_3$ from strain-stress curves, such as numerical differentiation [5,19], fitting the strain-stress relationship [7]. Another is the Lagrangian strain modes, and different combinations has been developed [5,7,13]. To address the two challenges, the essential is to develop some criteria and strategies for adjusting above two things.

For the challenge of efficiency, we proposed that each strain mode should provide as many independent equations as possible, which will sharply reduce the required number of strain modes and then reduce the number of DFT-based calculations. Taking CI (Laue group) symmetry as an example, one strain mode (see **Table I**) is enough to get full SOECs and TOECs, which is as simple as calculating SOECs alone.

TABLE I A possible effective strain mode and corresponding coefficients for calculation of TOECs in CI symmetry. This strain mode is used for the calculation of TOECs of CI symmetry in current work if not specified.

| No. | $A_2$ | $A_3$ | $t$ | Strain |
|---|---|---|---|---|
| 1 | $0.5C_{11}-2.5C_{12}$ | $0.25C_{111}+1.75C_{112}+2C_{123}+C_{144}+C_{155}$ | $t_1$ | |
| 2 | $-2C_{11}$ | $4C_{111}+0.5C_{112}-0.5C_{123}+C_{144}+C_{155}$ | $t_2$ | $\begin{bmatrix} 0.5\eta & 0 & 0.5\eta \\ 0 & -2\eta & 0.5\eta \\ 0.5\eta & 0.5\eta & -0.5\eta \end{bmatrix}$ |
| 3 | $-0.5C_{11}-1.5C_{12}$ | $0.25C_{111}+5.75C_{112}-2C_{123}+2C_{155}$ | $t_3$ | |
| 4 | $C_{44}$ | $C_{144}-5C_{155}$ | $t_4$ | |
| 5 | $C_{44}$ | $-4C_{144}$ | $t_5$ | |
| 6 | 0 | $2C_{456}$ | $t_6$ | |

Under this framework, **TABLE II** shows the required number of strain modes of TOECs for all crystal classes, and the number of strain modes of the present method is 1/6 to 1/4 of conventional strain-energy method [14]. Compared with previous strain-stress method, the number of strain modes is 1/3 for cubic [7,12,13] and ~1/2 for triclinic symmetry [5]. A possible set of strain modes for calculating TOECs of arbitrary symmetry are listed in our joint submission [27].

TABLE II. The number of independent TOECs and necessary number of Lagrangian strain modes for arbitrary symmetry.

| Laue group | CI | CII | HI | HII | TI | TII | RI | RII | O | M | N |
|---|---|---|---|---|---|---|---|---|---|---|---|
| Number of independent TOECs | 6 | 8 | 10 | 12 | 12 | 16 | 14 | 20 | 20 | 32 | 56 |
| Number of strain modes | 1 | 2 | 2 | 2 | 2 | 3 | 3 | 4 | 4 | 6 | 10 |

**Fig. 1** shows the relative time of different methods to calculate TOECs of diamond. In the numerical-differentiation-based method, the time cost in current method is 1/5~1/3 of Cao's method [19] in the same accuracy level, and 4th order accuracy in current method(SSM-ND-acc4) is about 1/2 of Cao's 2nd order accuracy (SEP-ND-acc2). For the fitting-based method, our scheme is faster than traditional SEM by about 3 times, and 1.5 times for LSUS method (even without taking the time for pressure dependent of SOECs into consideration).

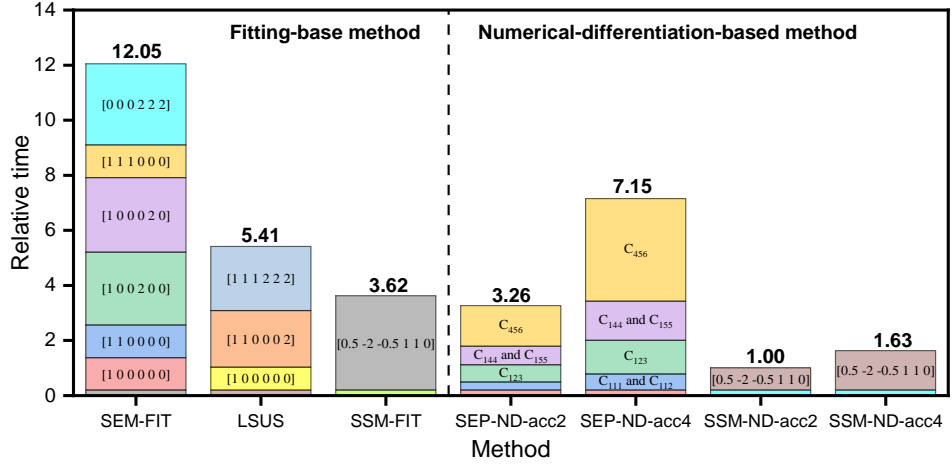

**Fig. 1** Cost of various methods of calculating TOECs of diamond. The value is normalized according to the time of SSM-ND-acc2 method. The label in each sub-block is the time spent in each strain mode (SEM, SSM or LSU) or elastic constants (SEP-ND). The strain range for all methods is -6% to 6%, except for LSUS which is -8% to 0. 9 uniform points were calculated for fitting-based method (SEM, SSM-FIT and LSUS), and the number of points is variable in numerical differentiation method (SSM-ND and SEP-ND) method according to the elastic constants and accuracy (ref [27] for details). For LSUS, only the time of uniaxial strain is calculated here. **SSM-FIT**: current fitting-based strain-stress method; **SSM-ND-acc2(4)**: current numerical-differentiation-based strain-stress method with 2nd (4th) order accuracy; **SEM-FIT**: strain-energy method; **SEP-ND-acc2(4)**: the separation of elastic constants method based on ref. [19] with 2nd (4th) order accuracy; **LSUS**: the longitudinal stress-uniaxial strain approach [7]. The bottom item in each method represents structural relaxation.

Since the possible combination of strain modes is countless, we need a strategy to determine the optimal ones. From **Eq.1**, we note that the coefficient matrix (**B** composed of $B_3^{\beta,\gamma}$ in **Eq.1**) of the linear equations is determined by the strain modes only, and the upper boundary of error on the TOECs can be evaluated by the condition number($\|\mathbf{B}\|\cdot\|\mathbf{B}^{-1}\|$) of coefficient matrix when introducing perturbation ($\partial A_3$) into the linear equation system (as **Eq.2**).

$$\frac{\partial C}{C} \leq \|\mathbf{B}\|\cdot\|\mathbf{B}^{-1}\|\frac{\partial A_3}{A_3} \qquad (2)$$

**Fig.2** shows the $C_{111}$ of diamond obtained by different strain modes with different condition numbers. The error increases with the increase of condition number. Because each TOEC is separated in SEP-ND, the condition number of SEP-ND equals to 1, which means that SEP-ND is the most robust method in the same accuracy. The condition number of the strain mode in **Table I** is 3.55, which is better than SEM (58.03) [14] and LSUS (9.73) [7] method. Using the condition number as an objective function, we optimized the strain mode for all classes of symmetry by particle swarm optimization [28] The best strain mode for CI symmetry is [0.29, -1.0, -0.41, 0.53, 0.73, 0], and corresponding condition number is about 2.94.

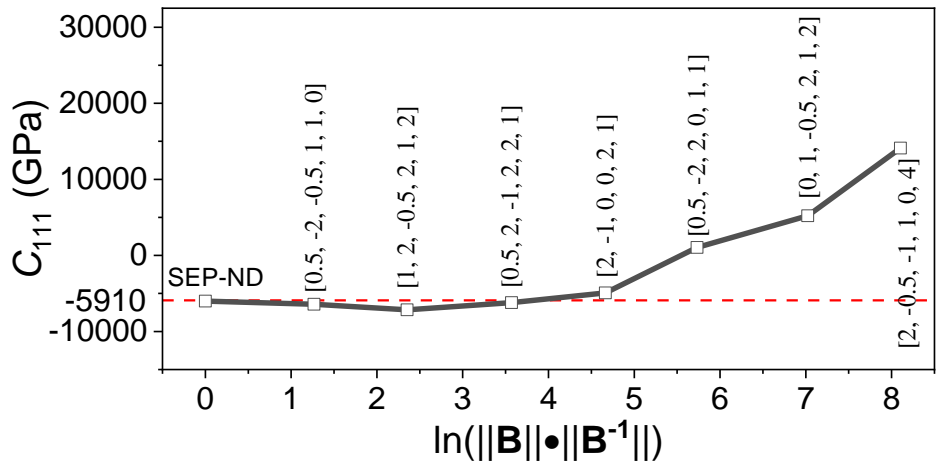

**Fig.2** The effect of strain mode on TOECs. The SSM-ND-acc2 is used for calculating $C_{111}$ and the strain amplitude is -6% to 6%. The red dash line is the recommended value of $C_{111}$ in current work.

Besides the condition number, the higher-order effect affects the accuracy of the method, which causes TOECs sensitive to the strain amplitude [5,14,29], and only a small strain plateau is found to get correct TOECs [5,29,30]. We proposed the use of higher-order polynomial (irrespective of the exact expression) in the fitting or higher-order-accuracy differential stencil for numerical-differentiation-based method in evaluating the value of $A_2$ and $A_3$ in **Eq.1**. **Fig.3** shows the SOECs and TOECs of diamond under different strain amplitudes fitted by higher-order polynomial. When fitting **Eq.1** with $n$=3 ($n$ is the highest elastic constants considered, hereinafter), the maximum strain works only for a small window for obtaining the correct TOECs (~2%-4%). Even worse, an error occurs in the SOECs in this range. At 10% maximum strain, fitting with $n$=3 induces 38GPa (483GPa) error on $C_{11}$ ($C_{111}$). However, when increasing $n$ to 5, one can get a robust result in a wide strain range (~3%-10% for TOECs and 0-10% for SOECs). In addition, as the condition number predicted, the SEP-ND is the most robust and SEP-ND-acc4 works well in a wide strain range (1%-10%) for both SOECs and TOECs. The error term of $n$=3 or second-order accuracy SEP-ND can be perfectly fitted by quadratic function of maximum strain, meaning that the fifth-order effect is required to get accurate TOECs of diamond.

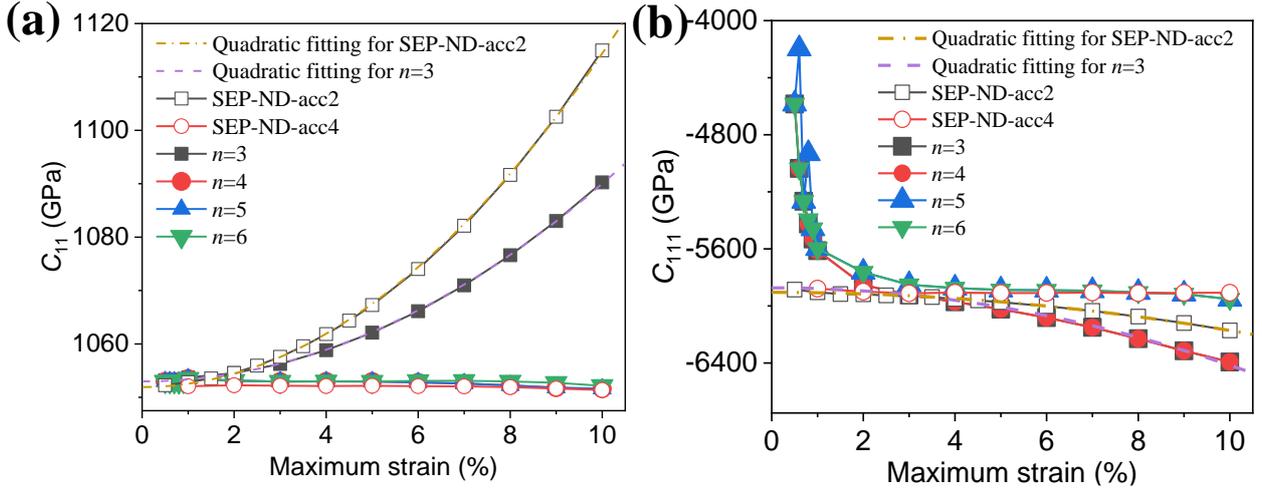

**Fig.3** The (a)$C_{11}$ and (b)$C_{111}$ of diamond obtained by weighted-fitting of **Eq.1** with different strain amplitudes when taking different higher-order effect into consideration. Here the strain ranges from -$\eta_{max}$ to $\eta_{max}$, 9 points are used in each fitting. SEP-ND-acc2 is the same with Cao's method [19], and it is expanded to 4$^{th}$ order accuracy in current work (SEP-ND-acc4), for detail equations, ref. [27].

For the case of symmetric strain, the fitting of order $n$ and $n+1$ with even (odd) $n$ gives the same value of SOECs (TOECs), as shown in **Fig.3a** and **3b**, which means that the fourth-order effect on TOECs is eliminated in symmetrical strain. However, instead of the symmetrical strain, the uniaxial compression is often used in experiments [18]. **Fig.4** shows the TOECs of diamond fitted by asymmetric strains. As it can be seen, a large variation occurs in the TOECs when $n$=3. For instance, when the central strain equals to 4%, $C_{111}$=-10502GPa (see **Fig.4a**), while the result by symmetrical strain is -6093GPa that is much close to the correct value (-5910GPa, suggested by current paper, ref. **Table III**). It demonstrates that the higher-order effect is non-negligible for diamond at uniaxial loading condition.

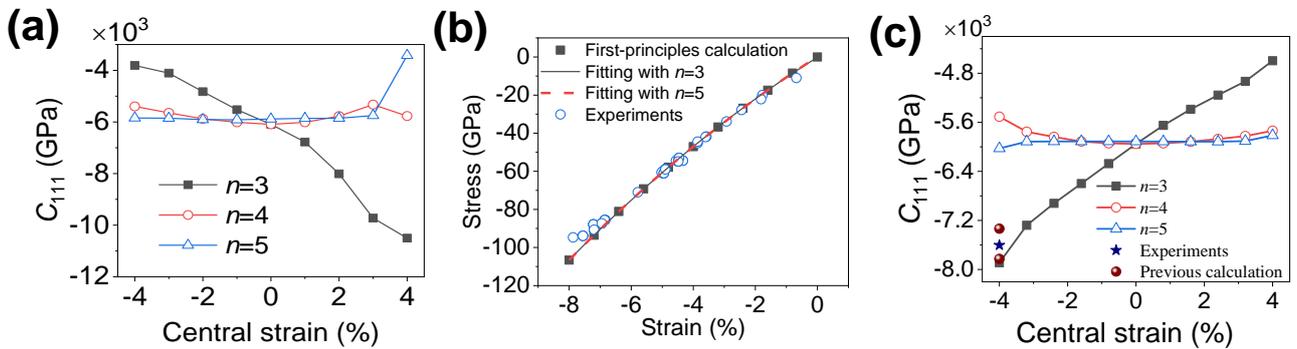

**Fig.4 (a)** The $C_{111}$ fitted by asymmetrical strain with different orders. **(b)** The calculated and fitted strain-stress curve along [100] direction under uniaxial compression and compared with experiments [13,31–33]. **(c)** The longitudinal elastic constants along [100] direction fitted by asymmetrical strain with different orders and compared with experiments [18] and previous calculations [7]. Note: the strain amplitude is 8%. 9 points are used for each fitting in (a) and 11 points for (b) and (c).

Based on the above discussion, we re-calculated the TOECs of diamond via Hmiel's method [7],

and the results are shown in **Fig.4** and **Table III**. Results for both $n=3$ and $n=5$ fit well with first-principles calculations and experiments [13,31–33] (**Fig.4b**), and the longitudinal elastic constants with n=3 agree well with previous calculations [7] and experiment [18] (**Table III**). However, it is evident that a large deviation occurs for $C_{111}$ when $n=3$ and central strain equals to -4% (**Fig.4c**). **Table III** show the comparison of SOECs and TOECs between current work and previous calculations and experiments. The results from the LSUS-A (LSUS with asymmetrical strain) method with $n=3$ agree well with experiments [21] and Hmiel's calculations [7], but with large deviation with strain-energy [7,20] or other strain-stress [19] method. This was mainly ascribed to the different frequency of each TOEC in the fitting equations in Hmiel's work. However, as $n$ increase to 5, the difference among different fitting methods (SEM [7,20], SEP-ND [19], LSUS-A, LSUS-S, SSM-FIT and SSM-ND) becomes quite small, suggesting that a large error of TOECs in diamond is due to the omission of higher-order terms in calculations which are present in shock wave compression experiments [18] and Hmiel's calculations [7]. Because of the minimum in condition number, the result by SEP-ND-acc4 (last column in **Table III**) is recommended as the accurate SOECs and TOECs of diamond.

**Table III.** The SOECs and TOECs of diamond by different methods. The strain range used in current work is [-8%, 0] for LSUS-A, [-8%, 8%] for LSUS-S and SSM. The SOECs and pressure dependence of SOECs for LSUS-A (-S) are taken from Ref. [7]. Unit in GPa.

| | Experiment [21] | Telichko's calc [20] SEM | Hmiel's calc [7] | | Cao's calc [19] SEP-ND | Present work | | | | | | |
|---|---|---|---|---|---|---|---|---|---|---|---|---|
| | | | LSUS | SEM | | LSUS-A[c] ($n=3$) | LSUS-A[c] ($n=5$) | LSUS-S[d] ($n=5$) | SSM-FIT ($n=3$) | SSM-FIT ($n=5$) | SSM-ND-acc4 | SEP-ND-acc4 |
| $C_{11}$ | 1079±5[a] | 1051 | 1033[b] | 1054 | 1037 | 1054 | 1054 | 1054 | 1077 | 1052 | 1051 | **1052** |
| $C_{12}$ | 124±5[a] | 125 | 136[b] | 129 | 120 | 129 | 129 | 129 | 127 | 119 | 119 | **119** |
| $C_{44}$ | 578±2[a] | 560 | 528[b] | 559 | 552 | 559 | 559 | 559 | 576 | 562 | 561 | **562** |
| $C_{111}$ | -7600±600 | -7611 | -7515 | -6026 | -5876 | -7890 | -6026 | -5907 | -6230 | -5905 | -5906 | **-5910** |
| $C_{112}$ | -1270±570 | -1637 | -845 | -1643 | -1593 | -689 | -1620 | -1680 | -1619 | -1571 | -1594 | **-1593** |
| $C_{123}$ | -330±920 | 604 | -960 | 606 | 618 | -1297 | 566 | 685 | 738 | 665 | 637 | **642** |
| $C_{144}$ | 2390±850 | -199 | 2693 | -200 | -197 | 3359 | 5 | -304 | -181 | -201 | -189 | **-198** |
| $C_{155}$ | -4100±380 | -2799 | -4223 | -2817 | -2739 | -4576 | -2899 | -2745 | -2973 | -2768 | -2766 | **-2779** |
| $C_{456}$ | -2890±750 | -1148 | -2870 | -1168 | -1111 | -2961 | -1278 | -1072 | -1218 | -1148 | -1148 | **-1157** |

[a]Reference [34].
[b]Calculated according to ref. [7]
[c,d]LSUS-A (-S): LSUS with asymmetrical (symmetrical) strain and a typo of **Eq.(A11)** in ref [7] is corrected.

In conclusion, this work offers a deep insight into understanding the relationship between TOECs and strain, including the mode, type (symmetrical or not) and amplitude of strains, and presents an efficient and accurate approach for the calculation of TOECs. The new approach includes: (1) development of the independent coefficient matrix with fewer strain modes for efficiency and smaller condition number of the coefficient matrix for robustness; and (2) a fitting strategy with higher-order terms to eliminate the sensitivity of TOECs on strain amplitude. It is demonstrated that the differences on the TOECs of diamond among experiments and computations in the literature can be understood and resolved by the proposed approach.

This work was supported by the China Scholarship Council and the China Postdoctoral Science Foundation also funded this project. The author also would thank Dr. Jinglian Du for useful discussion.